\documentclass[twocolumn,prb,aps,superbib,tightenlines,floatfix]{revtex4}

\usepackage{amsfonts}
\usepackage{amsmath}
\usepackage{amssymb}

\usepackage{graphicx}

\begin{document}

\title{Approach to steady-state transport in nanoscale conductors}

\author{Neil Bushong, Na Sai, and Massimiliano Di Ventra}

\affiliation{Department of Physics, University of California, San
Diego, La Jolla, CA 92093-0319}

\begin{abstract}
  We show, using a tight-binding model and time-dependent
  density-functional theory, that a quasi-steady state current can be
  established dynamically in a finite nanoscale junction without any
  inelastic effects. This is simply due to the geometrical
  constriction experienced by the electron wavepackets as they
  propagate through the junction. We also show that in this closed
  non-equilibrium system two local electron occupation functions can
  be defined on each side of the nanojunction which approach Fermi
  distributions with increasing number of atoms in the electrodes. The
  resultant conductance and current-voltage characteristics at
  quasi-steady state are in agreement with those calculated within the
  static scattering approach.
\end{abstract}

\maketitle

The static scattering approach has been extensively used to treat
steady-state transport in mesoscopic and nanoscopic conductors. The
approach, as originally introduced by Landauer, treats the sample as a
scatterer between two leads, which are connected adiabatically to two
infinite electron reservoirs at different local electrochemical
potentials.~\cite{landauer57,buttiker85} The reservoirs are just
conceptual constructs which enable one to map the non-equilibrium
transport problem onto a static scattering
one.~\cite{imry86,landauer87} However, the ensuing steady state may
not necessarily be the ``true'' steady state that is reached
dynamically when a battery discharges across the sample. In addition,
the static picture says nothing about the dynamical onset of steady
states, their microscopic nature, or their dependence on initial
conditions. These issues are particularly relevant in nanoscale
structures where some of the assumptions of the static approach may
hide important physical properties pertaining to the true charge
dynamics.

In this paper we employ an alternative picture of transport in
nanoscale systems in which we abandon the infinite reservoirs invoked
by Landauer. Instead, as recently suggested by Di Ventra and
Todorov~\cite{diventra04}, we consider the current that flows during
the discharge of two large but {\it finite} oppositely-charged
electrodes connected by a nanojunction. Unlike the static, open
boundary approach, the present approach permits one to describe the
current within a microcanonical formalism where both energy and
particle numbers are conserved quantities. In addition, due to the
finite and isolated nature of the system, it can be
demonstrated~\cite{diventra04} that the {\it total} current flowing
from one electrode to the other can be calculated {\it exactly} using
time-dependent density-functional theory (TDDFT)~\cite{runge84}
provided that one knows the exact functional, regardless of whether
the system reaches a steady state or not.

\begin{figure}
\includegraphics[width=3.25in]{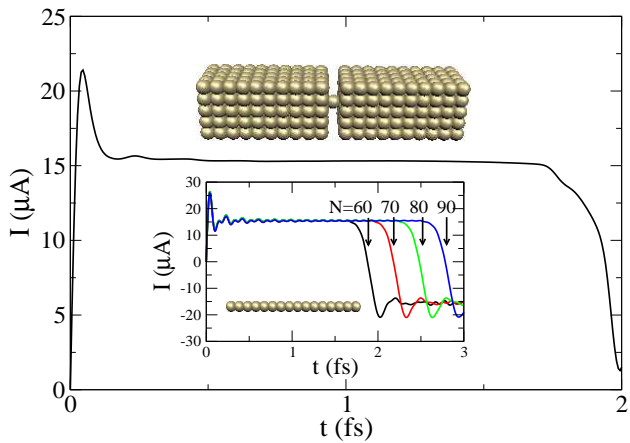}
\caption{Current passing through the junction as a function of time
  for a 3D nanojunction (schematic is shown) calculated with a
  non-interacting TB model. Each of the electrodes consists of
  $5\times 5\times30$ atoms arranged in a simple cubic geometry. The
  inset shows the corresponding current for a linear chain of N=60,
  70, 80 and 90 gold atoms. In both cases $E_B=0.2$ eV (see text).}
\label{fig:TBCurrent}
\end{figure}

We find that a quasi-steady state current, though lasting only for a
limited period of time, can be established in the neighborhood of the
nanojunction without any dissipation. This is simply due to the change
in the spread of momentum of wavepackets as they move into a
nanojunction and adapt to the given junction geometry. This effect
occurs roughly in a time $\Delta t \sim \hbar /\Delta E$, where
$\Delta E$ is the typical energy spacing of lateral modes in the
junction. For a nanojunction of width $w$, $\Delta E\sim
\pi^{2}\hbar^{2}/m_{{\rm e}}w^{2}$ and $\Delta t \sim m_{{\rm
    e}}w^{2}/\pi^{2}\hbar$. If $w$ = 1 nm, $\Delta t$ is of the order
of 1 fs, i.e., orders of magnitude smaller than typical
electron-electron or electron-phonon scattering times.~\cite{ash} We
indeed focus on the electron dynamics after the quasi-steady state has
been established and make a connection between this dynamical picture
and the Landauer's static approach. To this end, we consider a finite
three-dimensional (3D) model gold nanojunction and a finite
quasi-one-dimensional (1D) gold wire (see schematics in
Fig.~\ref{fig:TBCurrent}). These are the simplest structures for which
the quantized conductance and current-voltage characteristics have
been computed using the static scattering
approach~\cite{REFTodorovStatic,lagerqvist04} and have been measured
experimentally for similar gold quantum point
contacts.~\cite{agrait03} Recently, Horsfield {\it et al.} have shown
that a steady current is generated in a similar finite atomic
chain.~\cite{horsfield04p8251} Nevertheless the question of whether a
steady state can be reached without including any electron-ion
interactions remains unanswered in their work.~\cite{Private} In
addition to answering this question, we show that one can define two
local electron occupation functions on each side of the nanojunction.
These are shifted in energy by an amount which can be interpreted, in
the limit of large electrodes, as the ``bias'' of the corresponding
open system. These functions depart from the equilibrium Fermi
distributions by an amount which decreases with increasing electrode
size. This verifies Landauer's hypothesis that ``geometrical
dilution'' of wavefunctions is the most important aspect of a
reservoir.~\cite{landauer89} However, contrary to previous
conclusions,~\cite{landauer92,payne89} we show that finite but long
one-dimensional leads do not need to widen to constitute good
``reservoirs'', as long as one considers the electron dynamics in the
junction before the electrons reach the edge of the system.

We now begin our study by using a simple time-dependent tight-binding
(TB) model for noninteracting electrons where Coulomb interactions and
correlation effects are absent. Later, we treat the problem using a
fully self-consistent TDDFT approach in the adiabatic local density
approximation (ALDA).~\cite{ALDA}

Consider the $N$-site single-orbital TB Hamiltonian
\begin{equation}
\label{eq:TBH}
H^\mathrm{TB} = \sum_{i=1}^N \epsilon_i |r_i \rangle \langle r_i|
+ \mathfrak{t} \sum^N_i |r_i \rangle \langle r_{i+1}| + {\rm H.c.},
\end{equation}
where there is one orbital state $|r_i\rangle$ per atomic site with
energy $\epsilon_i$ and transfer matrix element $\mathfrak{t}$
connecting nearest-neighbor sites.~\cite{note-tb} We then prepare the
system such that half of the system has a deficiency of electrons, and
the other half has a surplus. This can be done by increasing
${\epsilon_i}$ of the sites on one side of the system by an energy
``barrier'' $E_B$. For the 1D wire (see inset of
Fig.~\ref{fig:TBCurrent}), the interface between the two regions
separated by the barrier defines the nanoscale ``junction''. Taking
this state as the initial state of the system, we then remove $E_B$,
and let the electrons propagate according to the time-dependent
Schr\"{o}dinger equation (TDSE) with the time-independent Hamiltonian
$H^{TB}$.~\cite{note-initial}

Due to the closed and finite nature of the system, the total current
can be calculated by time differentiating the charge accumulated on
one side of the system, i.e.,
\begin{equation}
I(t) = -e\frac{d}{dt}\sum_{n=1}^{N/2} \sum_{i=1}^{N_L}
\langle \psi^n(t) | r_i \rangle
\langle r_i | \psi^n(t) \rangle.
\label{eq:I}
\end{equation}
Here $\psi^n(t)$ are the occupied single-electron states that are
solution of the TDSE, and $N_L$ is the number of sites on the left of
the junction interface. Summation over spin degrees of freedom is
implied.

The onset of a quasi-steady state for a 3D nanojunction is shown in
Fig.~\ref{fig:TBCurrent}, where we plot the current Eq.~(\ref{eq:I})
as a function of time for $E_B=0.2$ eV. In the inset, we show that a
similar quasi-steady state current develops in 1D wires of different
lengths, where the initial time energy barrier forces electrons to
change the spread of electron momentum, and hence plays a role similar
to that of the geometric constriction in the 3D case. In all cases,
the current initially rises rapidly, but quickly settles in a
quasi-constant value $I_\mathrm{ss}$.\cite{note-discrete} In the 1D
structures, small oscillations are observed which decay in time. The
quasi-steady state lasts for a time $t_d$ during which the electron
waves propagate to the ends of the wire and back. The time $t_d$ is a
few femtoseconds for the considered cases, and can be made longer by
increasing the length of the wires (see Fig.~\ref{fig:TBCurrent}
inset). We have thus demonstrated numerically our initial conjecture:
in a closed and finite nanoscale system, a quasi-steady state current
with a finite lifetime can develop even in the {\it absence} of
dissipative effects. The steady state is a direct consequence of the
geometrical constriction experienced by the electron wavepackets as
they propagate through the junction. This is in contrast with the
conclusion of Ref.\cite{kurth}, where the establishment of a steady
state is attributed to a ``dephasing mechanism'' of the electrons
spreading in infinite electrodes.

\begin{figure}
\includegraphics[width=3.25in]{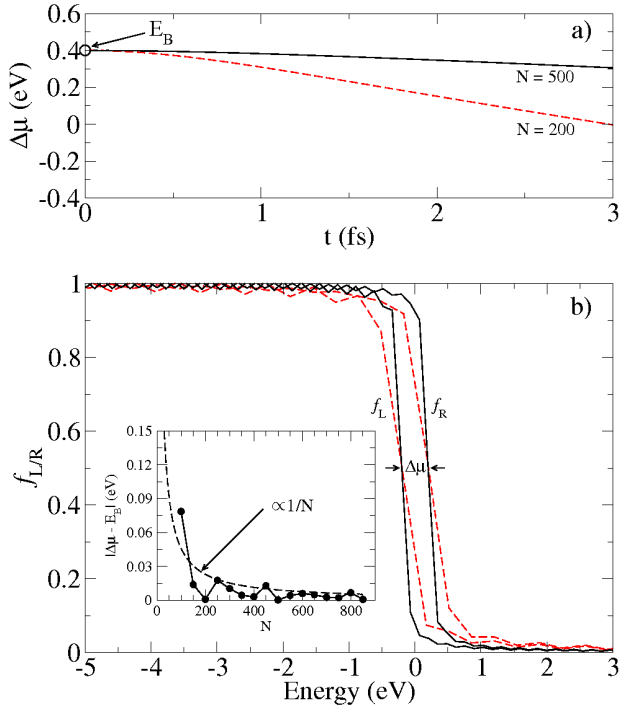}
\caption{Panel (a) illustrates $\Delta \mu(t)$ for a linear chain of
  $N=200$ (dashed line) and $N=500$ (solid line) atoms. Panel (b)
  shows their local occupation functions $f_L(E_j,t)$ and $f_R(E_j,t)$
  at a small time $t$ after the onset of current.  The inset shows the
  absolute difference between $\Delta \mu(t=0)$ and the initial energy
  barrier $E_B=0.4$ eV as a function of $N$. The dashed curve is
  proportional to the function $1/N$.}
\label{fig:TBDV}
\end{figure}

In order to calculate the conductance of this closed system, we need
to define a ``bias''. In this non-interacting electron problem, the
energy barrier $E_B$ seems a natural choice. However, that is not
completely satisfying as it relates to the initial conditions and not
to the electron dynamics. Let us instead define local occupation
numbers for electrons in the left and right regions of the system.
This concept is typically introduced as a starting point in the static
approach to transport. Here we would like to define it dynamically. We
then project the occupation for each eigenstate $|E_j \rangle$ of the
Hamiltonian $H^{TB}$, i.e, $f(E_j,t) = \sum_n |{ \langle E_j |
  \psi^n(t) \rangle |}^2$ onto the left- and right-hand side of the
system,
\begin{eqnarray}
\label{eq:locdist}
f(E_j,t) & = &
\sum_n \Big| \sum_{i \leq N/2} \langle E_j | r_i \rangle
\langle r_i | \psi^n(t) \rangle \Big|^2\nonumber\\
& &+\sum_n\Big| \sum_{i > N/2} \langle E_j | r_i \rangle
\langle r_i | \psi^n(t) \rangle \Big|^2 \nonumber \\
& & + \sum_n 2 \mathrm{Re}\bigg\{
\sum_{i \leq N/2} \langle E_j | r_i \rangle
\langle r_i | \psi^n(t) \rangle \nonumber \\
& & \sum_{i > N/2} \langle \psi^n(t) | r_i \rangle
\langle r_i | E_j \rangle \bigg\}.
\end{eqnarray}
We denote the first, second, and third term $f_L(E_j,t)$,
$f_R(E_j,t)$, and $f_C(E_j,t)$, respectively. The quantity
$f_C(E_j,t)$ is the sum of cross terms between the energy distribution
on the left and on the right region. Note also that because the set
$\{| E_j \rangle\}$ forms a complete orthonormal basis, $\sum_j
f_L(E_j,t) = n_L$, $\sum_j f_R(E_j,t) = N - n_L$, and $\sum_j
f_C(E_j,t) = 0$, where $n_L$ is the total charge left to the junction
at a given time. The quantities $f_L(E_j,t)$ and $f_R(E_j,t)$,
normalized to two electrons per state, are plotted in
Fig.~\ref{fig:TBDV}(b) for two wires of $N=200$ and $N=500$ atoms
immediately after the onset of the current.

One might naively think of these functions as broadened Fermi
distributions centered at different ``chemical potentials'', separated
by an energy $\Delta \mu(t)$. A closer examination, however, reveals
that they cannot be simply fitted to Fermi functions with just an
effective thermal broadening. Instead, the very functional form of
these non-equilibrium functions is different from a Fermi
distribution.  This is not surprising, as in this finite dynamical
system electrons spread on each side of the junction and are not in
any sort of local equilibrium in the electrodes. In addition, $\Delta
\mu(t)$ decreases with time because of the transport of electrons from
one side of the system to the other (see Fig.~\ref{fig:TBDV}(a)).
However, as long as we evaluate $f_L(E_j,t)$ and $f_R(E_j,t)$ at times
much less than $t_d$, these functions approach two zero-temperature
Fermi distribution functions centered at two different energies
$\mu_L$ and $\mu_R$, and $f_C(E_j,t)$ tends to zero for every $E_j$
with increasing number of sites $N$ in the electrodes.  These energies
can be interpreted as two local ``chemical potentials'' on the left
and right side of the system, with $\Delta \mu=\mu_R-\mu_L$
approaching the initial energy barrier $E_B$ with increasing $N$.

This asymptotic behavior as $N\rightarrow\infty$ is illustrated in
Fig.~\ref{fig:TBDV}(b) where $f_L(E_j,t)$ and $f_R(E_j,t)$ at a small
$t$ ($0 < t \ll t_d$) are plotted for different values of $N$. The
inset shows the absolute difference between $\Delta \mu(t=0)$ and
$E_B$ as a function of $N$. The difference scales with $N$ in the same
way as does the separation of eigenstates of $H^{TB}$ close to $\mu_L$
and $\mu_R$. In three dimensions $N=N_x\times N_y \times N_z$, and it
is then easy to prove, in this simple TB model, that $\Delta \mu(t=0)$
approaches (albeit ``nonvariationally'') $E_B$ as
$\frac{1}{N_x}+\frac{1}{N_y}+\frac{1}{N_z}$ with increasing number of
atoms in the three different directions (see inset of
Fig.~\ref{fig:TBDV}(b) for the 1D case). It is therefore evident that
with increasing $N$, local equilibrium distributions can be
effectively achieved in the two electrodes without inelastic effects:
the electron waves moving into these regions are geometrically
``diluted'' in a practically infinite region of space and therefore do
not ``disturb'' the local electron occupation. This is the equivalent
of Landauer's definition of reservoirs.~\cite{landauer92} Our results,
however, show that this definition can be extended to one-dimensional
electrodes as well.

\begin{figure}
\includegraphics[angle=270, width=3.25in]{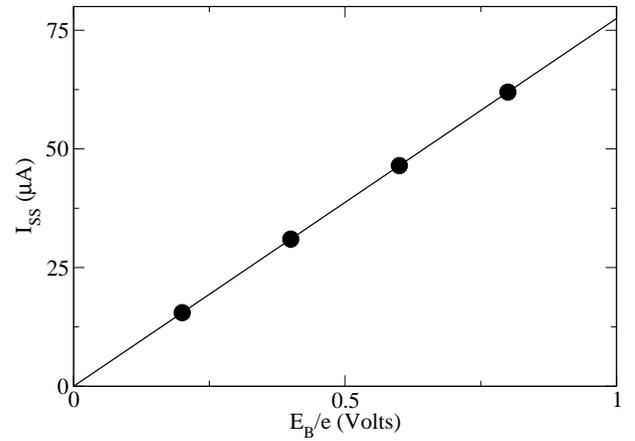}%
\caption{Current-voltage (I-V) characteristics at quasi-steady state
  of a finite 1D gold wire obtained using the TB approach. The
  corresponding conductance is $1.0\;G_0$. The TDDFT calculation
  yields a conductance of $0.99 \pm 0.03\; G_0$.}
\label{fig:IV}
\end{figure}

All this discussion allows us to define a conductance in this closed
system in terms of the current at steady state $I_{ss}$ and the value
of $\Delta \mu(t)/e$ for $N\rightarrow \infty$. The former converges
very fast with increasing number of atoms, whereas the latter is the
desired ``bias'' which, in turn, is simply the potential ``barrier''
$E_B/e$ at $t=0$. The current $I_{ss}$ as a function of $E_B/e$ is
plotted in Fig.~\ref{fig:IV}. The corresponding differential
conductance is about $1.0\;G_0$ ($G_0=2e^2/h$) at all voltages, in
good agreement with values obtained from the static
approach~\cite{REFTodorovStatic,lagerqvist04} and experimental
observations for similar systems.~\cite{agrait03}

\begin{figure}
\includegraphics[width=3.25in]{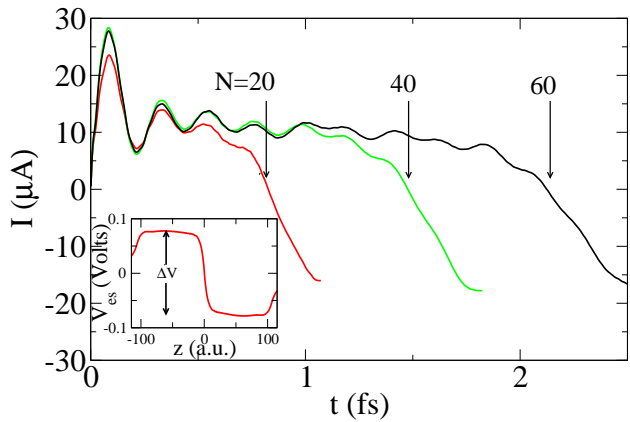}
\caption{Current in the middle of the junction as a function of time
  for a linear chain of N= 20, 40, 60 atoms calculated using
  TDDFT-ALDA. The inset shows the electrostatic potential drop along
  the wire at $t=0$ for a wire of 40 atoms.}
\label{fig:DFTCurrent}
\end{figure}

Finally, we study the onset of quasi-steady states in the presence of
electron interactions that we describe at the mean-field level. We
illustrate this point using TDDFT within the ALDA~\cite{socorro} for
1D wires. The corresponding current is plotted in
Fig.~\ref{fig:DFTCurrent} for different lengths of a finite chain of
gold atoms kept at a fixed distance of
2.8\AA$\;$apart.~\cite{note-tddft} The lifetime of the quasi-steady
state is short due to the limited system size but clearly increases
with increasing length of the wire. What is more interesting, however,
is the time for the quasi-steady state to set in. The initial
transient time is found to be less than 1 fs, consistent with our
original estimate.

The single-particle Kohn-Sham states~\cite{kohn65,runge84} have no
explicit physical meaning so that the interpretation of the
corresponding functions in Eq.~(\ref{eq:locdist}) is less clear. On
the other hand, the charge density, and thus, the electrostatic
potential, are well defined quantities. We therefore define the
conductance in this closed system in terms of the electrostatic
potential drop between two points inside each
electrode.~\cite{diventra04} As in the case of $\Delta \mu$, the
potential drop converges to the $t=0$ value (plotted in the inset of
Fig.~\ref{fig:DFTCurrent}) with increasing number of
atoms.~\cite{note-static} The corresponding differential conductance
is about $0.99\pm0.03 \; G_0$, where the average value has been
determined from the current in the wire with $N=60$ atoms at
$t=$1fs.~\cite{note-correction} It is worth pointing out that when the
hopping parameter in the tight-binding calculation is chosen to match
the drop-off time $t_d$ in the TDDFT calculation for the same number
of atoms, the initial transient time during which the quasi-steady
state establishes itself in the tight-binding calculation is also less
than 1 fs. This observation reinforces the notion that the geometric
constriction effect is present irrespective of the inclusion of
mean-field interactions.

Finally, one can obtain an order-of-magnitude estimate of the
electrode size necessary to observe the quasi-steady state. The
drop-off time $t_d$ is roughly given by the time it takes for an
electron to travel at the Fermi velocity of the underlying lattice
along the length of the electrode and back, i.e. $t_d \approx L /
v_f$, where $L/2$ is the linear length of one of the electrodes.
Therefore, one should observe a quasi-steady state if the time
necessary for the quasi-steady state to be established $\Delta t$ is
less than $t_d$, i.e. $L>v_f m_{{\rm e}}w^{2}/\pi^{2}\hbar$. The
results illustrated in Figs.~\ref{fig:TBCurrent}
and~\ref{fig:DFTCurrent} are consistent with these crude estimates and
show that a relatively small number of atoms is necessary to represent
the electrodes, thus making the present approach a practical
alternative to standard open-boundary calculations of transport.

In conclusion, we have shown numerically that a quasi-steady state can
be achieved in a nanoscale system without dissipative effects, simply
owing to the geometrical constriction experienced by electron
wavepackets as they approach the nanojunction. We have also provided a
practical scheme for dynamical conductance calculations in finite
nanoscale systems that sheds new light on the assumptions of the
standard static approach to steady-state conduction. The approach is
also suited to study relatively unexplored effects such as transient
phenomena, time-dependent charge disturbances, uniqueness of steady
states and their dependence on initial conditions.

We thank Ryan Hatcher for providing help with the time-dependent
calculations. We acknowledge financial support from the Department of
Energy (DE-FG02-05ER46204).



\begin{references}
  
\bibitem{landauer57} Landauer, R. {\it IBM J. Res. Dev.} {\bf 1957},
  {\it 1}, 223.
  
\bibitem{buttiker85} B{\"u}ttiker, M.; Imry, Y.; Landauer, R.; Pinhas,
  S. {\it Phys. Rev. B} {\bf 1985}, {\it 31}, 6207.
  
\bibitem{imry86} Imry, Y. In {\it Directions in Condensed Matter
    Physics}; Grinstein, G.; Mazenko, G.; World Scientific: Singapore,
  1986; pg 101.
  
\bibitem{landauer87} Landauer R. {\it Z. Phys. B} {\bf 1987}, {\it
    68}, 217.
  
\bibitem{diventra04} Di Ventra, M.; Todorov, T.N. {\it J. Phys. Cond.
    Matt.} {\bf 2004}, {\it 16}, 8025.
  
\bibitem{runge84} Runge, E.; Gross, E.K.U. {\it Phys. Rev. Lett.}
  {\bf 1984}, {\it 52}, 997.
  
\bibitem{ash} Ashcroft, N. W.; Mermin, N.D. {\it Solid State Physics;}
  Sounders College Publishing: Philadelphia, PA, 1976.
  
\bibitem{REFTodorovStatic} See, e.g., Todorov, T. N. {\it J. Phys.
    Cond. Matt.} {\bf 2002}, {\it 14}, 3049.
  
\bibitem{lagerqvist04} Lagerqvist, J.; Chen, Y.-C.; Di Ventra, M.
  {\it Nanotechnology} {\bf 2004}, {\it 15}, S459.
  
\bibitem{agrait03} Agra\"it, N.; Yeyati, A. L.; van Ruitenbeek, J. M.
  {\it Phys. Rep.} {\bf 2003}, {\it 377} 81.
  
\bibitem{horsfield04p8251} Horsfield, A. P.; Bowler, D. R.; Fisher, A.
  J.; Todorov, T. N.; S{\'a}nchez, C. G. {\it J. Phys. Cond. Matt.}
  {\bf 2004}, {\it 16}, 8251.
  
\bibitem{Private} A.P. Horsfield and T. N. Todorov have communicated
  to us that they also find steady currents in the absence of
  inelastic effects but they have never reported their results in any
  publication (private communication).
  
\bibitem{landauer89} Landauer, R. {\it J. Phys. Cond. Matt.} {\bf
    1989}, {\it 1}, 8099.
  
\bibitem{landauer92} Landauer, R. {\it Physica Scipta} {\bf 1992} {\it
    T42}, 110.
  
\bibitem{payne89} Payne, M. C.; {\it J. Phys. Cond. Matt.} {\bf 1989}
  {\it 1}, 4931.
  
\bibitem{ALDA} Zangwill, A.; Soven, P. {\it Phys. Rev. A} {\bf 1980}
  {\it 21}, {\it 1561}.
  
\bibitem{note-tb} The matrix element $\mathfrak{t}=-11$ eV is chosen
  so that the time scales of the TB calculation are comparable to
  those in the TDDFT calculation for the same number of atoms in the
  system.
  
\bibitem{note-initial} This is just one of the many (essentially
  infinite) initial conditions one can choose to initiate current
  flow. However, note that some initial conditions may not lead to a
  quasi-steady state. The question of the dependence of steady states
  on initial conditions has been addressed in
  Refs.~\onlinecite{diventra04} and \cite{kurth}, and will be analyzed
  in more detail in a future publication.
  
\bibitem{note-discrete} Note that, in the 3D case, a quasi-steady
  state can be established only if the typical energy spacing of
  lateral modes of the electrodes is much smaller than the
  corresponding one in the junction.
  
\bibitem{kurth} Kurth S.; Stefanucci, G.; Almbladh, C.-O.; Rubio, A.;
  Gross, E. K. U. {\it Phys. Rev. B} {\bf 2005}, {\it 72}, 035308.
  
\bibitem{socorro} The calculations reported here have been done using
  the {\tt socorro} package
  (http://dft.sandia.gov/Socorro/mainpage.html), adapted by Ryan
  Hatcher to perform time-dependent calculations.
  
\bibitem{note-tddft} As in the TB case, we construct the initial state
  of the system such that the left-hand side of the gold chain has a
  deficiency of charge, and the right-hand side has a surplus; we use
  a step-like potential to create this imbalance at $t=0$. The current
  is similarly determined by differentiating in time the charge
  accumulating on the left side.
  
\bibitem{kohn65} Kohn, W.; Sham, L. J. {\it Phys. Rev.} {\bf 1965},
  {\it 140}, A1133.
  
\bibitem{note-static} Note that in the static formulation of transport
  there is a conceptual (albeit numerically small, see e.g. Di Ventra,
  M.; Lang, N. D. {\it Phys. Rev. B} \textbf{2002}, {\it 65}, 045402)
  distinction between the chemical potential difference and the
  electrostatic potential drop, the conductance being usually defined
  in terms of the former.
  
\bibitem{note-correction} We have recently shown that there exist
  dynamical corrections to the electron conductance obtained using the
  ALDA which exist even in the limit of zero frequency (Sai, N.;
  Zwolak, M.; Vignale, G.; Di Ventra, M. {\it Phys. Rev. Lett.} {\bf
    2005}, {\it 94}, 186810). These corrections depend on the gradient
  of the electron density and are therefore negligible for a gold
  quantum point contact. Combined with the theorem in
  Ref.~\onlinecite{diventra04} on the exact value of the TDDFT total
  current in a closed and finite system, this result shows that the
  TDDFT-ALDA current of a gold junction at steady state is very close
  to the exact many-body value.

\end{references}
\end{document}